\documentclass[12pt,a4paper]{article}

\usepackage[utf8]{inputenc}
\usepackage[T1]{fontenc}
\usepackage{amsmath,amssymb,amsfonts}
\usepackage{graphicx}
\usepackage[margin=1in]{geometry}
\usepackage{natbib}
\usepackage{hyperref}
\usepackage{booktabs}
\usepackage{float}
\usepackage{setspace}
\usepackage{enumitem}

\hypersetup{
    colorlinks=true,
    linkcolor=blue,
    citecolor=blue,
    urlcolor=blue
}

\title{Decentralization of Agenda-Setting Power and Domain-Selective Bridging:\\Algorithm Design Beyond the Echo Chamber Debate}

\author{Masahiro Fujita\\
Faculty of Sociology (Psychology Division), Kansai University\\
ORCID: \href{https://orcid.org/0000-0003-1681-8221}{0000-0003-1681-8221}}

\date{}

\begin{document}

\maketitle

\begin{abstract}
Echo chambers are an inevitable consequence of the human cognitive system being evolutionarily designed to prioritize processing of high-relevance information at the small-group scale, combined with algorithms that optimize engagement as their sole objective. Conventional prescriptions that normatively criticize echo chambers and demand individual behavioral change have low feasibility given these cognitive constraints. This paper constructs an Agenda Democratization Index (ADI) that quantifies the decentralization of agenda-setting power using four variables---barrier to entry, granularity, interactivity, and feedback resolution---and a Social Information Health (SIH) model that integrates ADI with the strength of bridging mechanisms. Based on this model, we propose ``domain-selective bridging,'' which incorporates not only engagement but also bridging into algorithmic scoring functions, optimizing the bridging weight for each information domain based on verifiability ($V$) and collective scope ($S$). An agent-based simulation comparing three algorithm designs---no bridging, uniform bridging, and domain-selective bridging---demonstrated that domain-selective bridging substantially outperforms uniform bridging on a joint efficiency measure (SIH $\times$ user satisfaction)---by a factor whose absolute value is sensitive to Model 1's near-zero user satisfaction, but whose direction and dominance ranking are robust--- improving information sharing in domains relevant to collective decision-making while maintaining user experience in hobby and lifestyle domains. This paper reframes the echo chamber debate from a normative opposition over ``whether to eliminate echo chambers'' to an engineering design problem of ``in which information domains, to what degree, and through what algorithm design should bridging be implemented.''
\end{abstract}


\section{Redistribution of Agenda-Setting Power}

In democratic societies, the power to determine ``what constitutes an important issue''---agenda-setting power---was concentrated throughout the twentieth century in a small number of nationwide broadcast media organizations. The classical agenda-setting theory formulated by \citet{mccombs1972agenda} empirically demonstrated that mass media effectively monopolized the power to determine ``what people think about.'' A limited set of organizations---newspaper companies, television networks, and wire services---decided what to cover and what to omit, and these decisions defined the scope of public discourse.

This monopoly depended on extremely high barriers to entry for information dissemination. In television, broadcast licensing systems restricted entry; in newspapers, large-scale printing equipment and distribution networks were required. It was virtually impossible for individuals or small groups to disseminate information at a national scale. The monopoly of agenda-setting power was sustained not by the excellence of media organizations but by the height of technological and institutional barriers to entry.

It is important to note that newspapers, unlike television, had a distinct hierarchical structure. In the United States, national papers such as the \textit{New York Times} and \textit{Washington Post}, major regional papers such as the \textit{Chicago Tribune} and \textit{Los Angeles Times}, and countless local papers coexisted \citep{starr2004creation}. In Japan, national dailies (Asahi, Yomiuri, Mainichi, etc.), regional bloc papers (Chunichi, Nishinippon, etc.), and prefectural papers coexisted in a similar hierarchy. This structure is consistent with the argument developed in this paper that information relevance decays with geographic distance. Local and bloc papers survived because they could provide information more closely aligned with readers' relevance structures, and the phenomenon whereby bloc papers hold stronger subscription bases than national papers in their regions can be understood as a natural consequence of this distance decay. Nevertheless, the hierarchical structure of newspapers did not result in a decentralization of agenda-setting power. Papers at every level remained unidirectional information dissemination media, with editorial control firmly held by newsrooms. Feedback from readers was limited to circulation figures and letters to the editor, yielding extremely low feedback resolution $F$. Moreover, national papers indirectly influenced local paper agendas through wire service content distribution \citep{shoemaker2014mediating}, meaning that even with a hierarchical structure, substantive decentralization of agenda-setting power was limited.

The first change to this monopoly structure came with the proliferation of cable television and satellite broadcasting in the 1980s and 1990s, which brought multi-channel options. \citet{prior2007post} characterized this period as ``post-broadcast democracy'' and showed that the increase in media choices made it possible for audiences to avoid news. However, cable television operated at the channel level, and each channel still retained organizational gatekeeping functions. Barriers to entry decreased but did not reach zero. Agenda-setting power merely shifted from a few large organizations to many mid-sized organizations, and its fundamental structure remained unchanged.

A qualitative rupture occurred with the proliferation of systems---exemplified by YouTube---that allow anyone to distribute and receive video content over the Internet. These systems are operated on an advertising-revenue model, meaning neither senders nor receivers bear any cost. This economic structure effectively eliminated barriers to entry. In addition, the simultaneous emergence of granularity at the level of individual videos, bidirectional participation by viewers (comments, sharing, response videos), and a dramatic increase in feedback resolution through real-time analytics created a qualitative break from the cable era. Government agencies and politicians now publish complete footage of press conferences and policy briefings, enabling citizens to access primary source material directly. Whereas previously only information filtered through the editorial processes of nationwide broadcast media reached citizens, it is now possible for citizens themselves to compare primary sources with edited broadcasts. Agenda-setting power is structurally decentralizing from a few nationwide broadcast media organizations to a multitude of senders.

The decentralization of agenda-setting power is fundamentally a positive change for democracy. The transition from a regime in which a small number of organizations unilaterally decide ``what the public needs to know'' to one in which pluralistic information senders compete should be described as a democratization of power. However, this change has frequently been discussed negatively as the problem of ``echo chambers'' \citep{sunstein2001republic,sunstein2017republic,pariser2011filter}---the argument that algorithms recommending content tailored to users' preferences cause people to encounter only information that reinforces their existing beliefs, deepening societal divisions.

In response to this problem, proposals have recently been made to shift algorithmic scoring functions from engagement-based ranking to bridging-based ranking. \citet{ovadya2023bridging} proposed the concept of ``bridging systems'' to promote mutual understanding and trust across divides, illustrating concrete implementations such as Community Notes and Polis. This proposal represents an important contribution in the direction of addressing divisions through algorithm design.

However, the bridging systems framework has three important limitations. First, it does not systematically develop the argument that bridging weights should differ across information domains, treating bridging necessity as essentially uniform. Echo chambers in hobby and lifestyle information and those in policy and security discussions should have qualitatively different bridging requirements. Second, it does not provide a cognitive foundation for why echo chambers form in the first place. It starts from division as a ``problem'' but lacks the perspective that echo chambers are an inevitable phenomenon arising from constraints of the human cognitive system. Third, it lacks a framework for quantitatively describing the process of agenda-setting power decentralization itself, making it unclear how bridging fits within media history.

This paper addresses these limitations to provide a theoretical foundation for the bridging systems discussion and to refine it practically. The starting point is the cognitive foundation of echo chambers. The human cognitive system has severe constraints on processing capacity, and within that finite capacity, it has been evolutionarily designed to prioritize information relevant to one's survival and reproduction \citep{sperber1995relevance}. In evolutionary environments, information relevant to survival circulated primarily within the small group of approximately 150 individuals with whom one lived daily---the social network size estimated by \citet{dunbar1992neocortex}. Given these cognitive constraints, the convergence of information processing scope to small-group scale is a natural phenomenon aligned with the design of the human cognitive system. The premise that all humanity should share the same agenda and process the same information is merely the result of mistaking for a universal norm an exceptional state produced by the historically peculiar institution of national mass media, which has a history of barely 150 years. Prescriptions that normatively criticize echo chambers and demand individual behavioral change have low feasibility given the limits of cognitive capacity.

The problem lies not in the decentralization of agenda-setting power itself but in the insufficiency of bridging mechanisms connecting different information spheres in the post-decentralization information environment. However, the need for bridging is not uniform across all information domains. Information verifiable through personal experience (local shop reviews, hobby knowledge, etc.) functions sufficiently within echo chambers. Bridging is truly needed only in information domains that are difficult for individuals to verify and that influence collective decision-making---policy, security, public health, and the like. Introducing this distinction enables the optimization of bridging weights in algorithmic scoring functions for each information domain.

The contributions of this paper are threefold. First, we propose the Agenda Democratization Index (ADI) as an indicator that continuously quantifies the decentralization of agenda-setting power. ADI comprises four variables---barrier to entry, granularity, interactivity, and feedback resolution---and provides a unified description of each stage of media history from the broadcast monopoly era to the YouTube era. Second, we construct a Social Information Health (SIH) model that integrates ADI with the strength of bridging mechanisms, providing a framework for independently evaluating the degree of agenda-setting power decentralization and the adequacy of bridging. Third, we extend the bridging systems discussion of \citet{ovadya2023bridging} and propose domain-selective bridging based on information verifiability and collective scope. Through agent-based simulation, we compare three algorithm designs---no bridging, uniform bridging, and domain-selective bridging---and demonstrate that domain-selective bridging improves social information health more efficiently than uniform bridging.

The paper is organized as follows. Section~\ref{sec:review} reviews agenda-setting research and the bridging systems discussion, identifying limitations of existing research. Section~\ref{sec:evolutionary} builds the theoretical foundation of evolutionary cognitive constraints and information relevance distance decay based on relevance theory. Section~\ref{sec:model} presents the mathematical formulation of the ADI-SIH model. Section~\ref{sec:simulation} conducts agent-based simulation for dynamic analysis and model comparison. Section~\ref{sec:discussion} discusses theoretical and policy implications along with limitations, and Section~\ref{sec:conclusion} concludes.

\section{The Echo Chamber Debate and the Current State of Agenda-Setting Research}
\label{sec:review}

\subsection{Classical Agenda-Setting Theory and Its Assumptions}

Agenda-setting theory was formulated by \citet{mccombs1972agenda} based on a voter survey during the 1968 U.S.\ presidential election. Their finding was that the ranking of issues covered by the mass media and the ranking of issues perceived as important by voters showed a high correlation. This discovery empirically supported the insight of \citet{cohen1963press} that the media may not be successful much of the time in telling people what to think, but is stunningly successful in telling its readers what to think about.

What this theory implicitly assumed was the scarcity of information dissemination channels. Under broadcast licensing systems, the number of television stations was severely restricted, and publishing a newspaper required large-scale printing and distribution infrastructure. Information recipients had no alternative but to passively consume information selected by a small number of media organizations. The monopoly of agenda-setting power was a product of these technological and institutional conditions.

Agenda-setting research subsequently evolved to the second level (attribute agenda-setting, whereby not only ``what to think about'' but also ``how to think about it'' is shaped) and the third level (network agenda-setting, whereby the associative structure among issues transfers from media to audiences) \citep{mccombs2014setting,guo2016power}. However, at all levels, the basic structure in which a small number of senders set the agenda for a large number of receivers was maintained. The transformation of agenda-setting in social media environments has been actively studied in recent years \citep{meraz2009elite,vargo2014network}. \citet{chadwick2013hybrid} proposed the concept of the hybrid media system, analyzing the dynamics of mutual influence between old and new media logics. \citet{williams2011after} systematically discussed the new information environment following the collapse of broadcast-era media institutions. However, the question of how agenda-setting theory itself should be reconstituted when the premise of channel scarcity collapses has not been sufficiently addressed even in these studies.

\subsection{The Echo Chamber and Filter Bubble Debate}

The most widely discussed consequence of the decentralization of agenda-setting power has been the problem of echo chambers and filter bubbles.

In \textit{Republic.com}, \citet{sunstein2001republic} discussed the democratic risks of the Internet enabling fully customized information environments for users---what he called the ``Daily Me.'' His argument was that if citizens encountered only information confirming their existing beliefs and lost opportunities for chance encounters, the foundations of democratic deliberation would be undermined. \citet{sunstein2017republic} updated this argument for the social media environment. \citet{pariser2011filter} widely popularized the concept of the ``filter bubble'' created by personalization algorithms of search engines and social networking services.

However, the empirical evidence for echo chambers is not uniform. \citet{gentzkow2011ideological} showed that the ideological segregation of online news consumption is significantly lower than the segregation of face-to-face interactions with neighbors, co-workers, or family members. This finding casts doubt on the premise that echo chambers are a phenomenon unique to online environments. Recent large-scale experiments including those by \citet{guess2023algorithms,guess2023reshares} and \citet{nyhan2023likeminded} have also suggested that the causal impact of social media algorithms on polarization is not as large as previously assumed. \citet{dubois2018echo} showed that echo chamber effects are substantially mitigated among individuals with high political interest and diverse media consumption, pointing to the possibility that the echo chamber hypothesis has been overgeneralized. The Facebook study by \citet{bakshy2015exposure} showed that users' own selective exposure was more restrictive of ideologically cross-cutting information exposure than algorithms. On the other hand, the experiment by \citet{bail2018exposure} showed that exposure to opposing political views could strengthen polarization, particularly among Republicans, suggesting that simply ``exposing people to diverse information'' does not solve the problem.

\citet{donkers2021dual} distinguished between epistemic echo chambers and ideological echo chambers, using an agent-based model to show that fundamentally different diversification strategies are needed to resolve each type. This finding suggests the limitations of applying a uniform prescription to the echo chamber problem.

What is missing throughout this entire debate is the question of the cognitive mechanism behind why echo chambers form. The majority of existing research analyzes echo chambers as consequences of algorithm design or user behavioral choices, without considering the fact that the human cognitive system is evolutionarily designed to limit the scope of information processing to small-group scale. The premise that echo chambers are a ``problem'' may itself be nothing more than an evaluation using the information environment of the national mass media era as an implicit baseline.

\subsection{The Bridging Systems Proposal and Its Limitations}

The most systematic proposal for addressing echo chambers in recent years is the bridging systems framework of \citet{ovadya2023bridging}. They defined ``bridging systems'' as systems that promote mutual understanding and trust across divides, presenting concrete examples and open problems across three domains: recommender systems, collective response systems, and human-facilitated deliberation.

Their central proposal is bridging-based ranking. In place of the current engagement-based ranking---algorithms that maximize engagement metrics such as user clicks, shares, and dwell time---they propose a system that preferentially displays content receiving positive responses from users with diverse perspectives. Community Notes (formerly Birdwatch) on Twitter (now X) is positioned as an emergent implementation of this approach.

While the bridging systems proposal is an important contribution in the practical direction of improving the information environment through algorithm design, it has the three limitations mentioned above. First is the problem of uniform bridging application. Although Ovadya and Thorburn explicitly present many issues as open problems, they do not systematically develop the argument that bridging weights should differ by information domain. The necessity and optimal methods of bridging should be fundamentally different between echo chambers in hobby communities and those in policy discussions, but this distinction is not theoretically grounded.

Second is the absence of a cognitive foundation. No cognitive science explanation is provided for why humans form echo chambers or why exposure to bridging information sometimes backfires \citep{bail2018exposure}. If bridging design does not account for cognitive constraints, it faces essentially the same difficulties as the Sunstein-style prescription of ``expose yourself to diverse information,'' which has low feasibility.

Third is the absence of media-historical positioning. The bridging systems discussion starts with the current social media environment as given, lacking a historical and structural analysis of how agenda-setting power has decentralized and what stage this decentralization has reached. Without a framework that quantitatively describes what changed and what did not at each stage---broadcast monopoly, cable era, YouTube era---it is impossible to theoretically derive when and under what conditions bridging systems become necessary.

\subsection{Positioning of This Paper}

The above review reveals three gaps in the existing literature.

First, no framework exists for capturing the process of agenda-setting power decentralization with a continuous quantitative index beyond qualitative description. Second, no argument exists that theoretically grounds the formation of echo chambers in human evolutionary cognitive constraints and, on that basis, differentiates the need for bridging by information domain. Third, no model exists for optimizing the bridging weight in algorithmic scoring functions according to the characteristics of information domains.

This paper addresses these three gaps by building a theoretical foundation based on evolutionary cognitive constraints in Section~\ref{sec:evolutionary}, quantifying the decentralization of agenda-setting power and the adequacy of bridging through the ADI-SIH model in Section~\ref{sec:model}, and verifying the effectiveness of domain-selective bridging through agent-based simulation in Section~\ref{sec:simulation}.

\section{Evolutionary Cognitive Constraints and Distance Decay of Information Relevance}
\label{sec:evolutionary}

\subsection{Finite Cognitive Resources and Selective Processing by Relevance}

The human cognitive system does not have the capacity to process all information present in the environment. Processing capacity is severely constrained in attention, working memory, and encoding into long-term memory alike. Under these constraints, the cognitive system processes input information selectively rather than indiscriminately. The criterion for selection is the relevance of information.

Relevance theory, as formulated by \citet{sperber1995relevance}, theoretically articulated that human cognition is designed to maximize relevance. In their framework, the relevance of information is determined by two factors. One is cognitive effects---the gains obtained from processing the information, such as revision of beliefs, expansion of knowledge, and improvement of inferences. The other is processing effort---the investment of cognitive resources required to understand and integrate the information. Relevance is higher when cognitive effects are large and lower when processing effort is large. The human cognitive system is tuned to prioritize the processing of information that maximizes relevance.

\subsection{The Structure of Information Relevance in Evolutionary Environments}

The evolutionary foundation of this cognitive design is clear \citep{tooby1992psychological}. In the environment in which humans evolved as a species---the greater part of the Pleistocene---information relevant to an individual's survival and reproduction arose from spatially and socially proximate environments. A nearby predator is more urgent than a distant one, and social dynamics within one's own group (who is trustworthy, who is cooperative, who is a defector) are more directly relevant to behavioral decisions than those of other groups. The cognitive effect of information is greater the closer the spatial and social distance to the source and smaller the more distant.

\citet{dunbar1992neocortex} estimated from the correlation between neocortex ratio and social group size in primates that the upper limit of social relationships that human cognition can stably maintain is approximately 150 individuals. This ``Dunbar's number'' also has significance from an information processing perspective. Information circulating within a social network of approximately 150 individuals has high cognitive effects for the individual---because it is information directly related to one's survival, reproduction, and social standing. Beyond this range, cognitive effects decline sharply and no longer justify the processing cost.

Therefore, the convergence of the human cognitive system's information processing scope to small-group scale is not a design defect but an adaptive allocation of finite cognitive resources. In evolutionary environments, expending cognitive resources on distant, irrelevant information was adaptively disadvantageous in the sense that it interfered with processing nearby, highly relevant information.

\subsection{A Distance Decay Model of Information Relevance}

From the above discussion, we can derive the principle that information relevance decays as a function of the distance between source and receiver. ``Distance'' here is a multidimensional concept encompassing not only spatial distance but also social distance (similarity of group membership, degree of shared interests) and topical distance (relevance to one's own life and concerns).

We formalize the relevance $R_{ij}$ that information item $i$ holds for individual $j$ as follows:

\begin{equation}
R_{ij} = \frac{E_{ij}}{P_{ij}} = \frac{\text{Cognitive Effects}}{\text{Processing Effort}}
\end{equation}

Cognitive effects $E_{ij}$ are a decreasing function of distance $d_{ij}$, and processing effort $P_{ij}$ is an increasing function of distance (more distant information requires more contextual supplementation, increasing processing cost). Relevance therefore decays at an accelerating rate with distance:

\begin{equation}
R_{ij} \propto \frac{1}{d_{ij}^{\lambda}} \quad (\lambda > 1)
\end{equation}

This distance decay varies in rate depending on the type of information. Here we introduce two variables.

\textbf{Verifiability ($V_k$):} The degree to which information in domain $k$ can be confirmed through an individual's direct experience. ``The reputation of a new restaurant in the neighborhood'' has high $V$; ``the long-term consequences of a foreign policy'' has low $V$.

\textbf{Collective Scope ($S_k$):} The scale of the number of people affected by decisions in information domain $k$. ``Individual dietary choices'' have low $S$; ``voting in an election'' has high $S$.

In information domains where $V$ is high, individuals can verify information themselves, so the cost of cognitive closure without bridging information from others is small. In domains where $S$ is high, collective decision-making requires integrating information from many people, so the cost of cognitive closure is large. The need for bridging is proportional to $S/V$.

\subsection{Structural Comparison of Pre-Modern Gossip Networks and SNS}

This theoretical framework enables a structural comparison of pre-modern and contemporary SNS information environments.

In pre-modern village societies, information circulated primarily through face-to-face gossip networks. These networks were based on geographic proximity, and the scope of information processing was naturally limited to within the range of Dunbar's number. This environment had two important characteristics. First, information verifiability was high. The content of gossip was in many cases confirmable through direct observation or experience (``the well in the neighboring village has dried up'' could be confirmed by visiting). Second, bridging mechanisms existed at low frequency. Traders, pilgrims, and movement of people through marriage functioned as bridges carrying information between different gossip networks. Such information may also have been conveyed through what \citet{granovetter1973strength} theorized as weak ties---that is, the structural property whereby non-intimate but connected relationships tend to occupy positions bridging dense groups served as a source of pathways through which information flowed between different groups. Survival-critical information---epidemics, external enemies, famine---propagated between groups, albeit at low frequency.

The contemporary SNS environment shares the structural characteristic of information processing scope converging to small-group scale with pre-modern environments, but differs fundamentally in its basis. SNS echo chambers are formed not by geographic proximity but by social proximity based on shared interests. This difference produces two important consequences.

First, information verifiability is substantially reduced. Much of the information circulating in interest-based echo chambers concerns abstract political topics that cannot be verified through individual direct experience. In pre-modern gossip networks, direct feedback from the environment placed an upper limit on cognitive closure, but this natural corrective mechanism is absent.

Second, the conditions for information movement between different information spheres change. In pre-modern times, physical movement associated with trade and marriage carried information from one group to another. This bridging was low-frequency but naturally occurring, and survival-relevant information propagated between groups. In SNS environments, where information environments are complete without physical movement, naturally occurring bridging is less likely to arise. On the other hand, intentional bridging design through algorithms has become technically possible.

The above analysis demonstrates both the error of viewing the contemporary information environment as a simple ``regression'' to pre-modern conditions and theoretically derives that the need for bridging is higher than in pre-modern times. Under the twin conditions of reduced verifiability and insufficient naturally occurring bridging, intentional design of bridging mechanisms---particularly in information domains where $V$ is low and $S$ is high---is essential.

\subsection{The Cognitive Mismatch of Nationwide Broadcast Media}

This theoretical framework explains why the monopoly of agenda-setting by nationwide broadcast media was cognitively unnatural. Nationwide media disseminate identical information to tens of millions of receivers. However, the majority of this information is spatially and socially distant for most receivers, with cognitive effects that do not justify the processing cost. A distant house fire, a traffic accident in another region, the movements of candidates in a different electoral district---these are pieces of information that do not influence receivers' behavior, and from the distance decay model of relevance, it is cognitively rational for them not to be processed.

During the era when nationwide media monopolistically set agendas, receivers were passively exposed to such low-relevance information. This may have been beneficial for democracy as ``chance encounters'' \citep{sunstein2001republic}, but from the perspective of cognitive resource allocation, it was inefficient. The norm that all humanity should process the same agenda may have functioned as an ideology justifying this inefficiency.

However, information domains exist in which nationwide media legitimately function. Large-scale disasters, security threats, nationwide institutional changes, and other information where $V$ is low and $S$ is extremely high---information difficult for individuals to verify but affecting collective decision-making by the entire citizenry---can be cognitively justified for wide-area distribution by nationwide media. The problem is that nationwide media have claimed agenda-setting power over all agendas beyond this limited legitimate domain.

The next section builds a mathematical model integrating the decentralization of agenda-setting power and bridging mechanisms on this theoretical foundation.

\section{Mathematical Formulation of the ADI-SIH Model}
\label{sec:model}

\subsection{The Agenda Democratization Index (ADI)}

With the cognitive constraints discussed in the previous section as premises, we construct a framework for quantitatively describing the decentralization of agenda-setting power. The Agenda Democratization Index (ADI) proposed in this section captures structural characteristics of the media environment through four variables and expresses the degree of decentralization continuously through their multiplicative combination.

The first variable is barrier to entry ($B$), representing the capital and institutional costs required for a new information sender to enter the market. This includes acquiring broadcast licenses, equipping printing facilities, and building distribution infrastructure; the higher $B$ is, the more the number of senders is restricted and the more agenda-setting power concentrates in a small number of organizations. The second variable is granularity ($G$), representing the fineness of the minimum unit of information dissemination. In broadcast television, the unit is a program; in cable television, a channel; in YouTube, an individual video. The higher $G$ is, the more senders can specialize in specific topics and interests, enabling more precise correspondence to receivers' relevance structures. The third variable is interactivity ($I$), representing the degree to which receivers can participate in the process of agenda formation. In environments like broadcast television where receivers only passively consume programmed information, $I$ is near zero, but in environments like YouTube where receivers can participate in agenda formation through comments, sharing, and response videos, $I$ takes a positive value. The fourth variable is feedback resolution ($F$), representing the granularity and speed at which receiver responses are returned to senders. Newspaper circulation figures are monthly aggregate indicators for the entire publication and $F$ is extremely low, but YouTube analytics provide second-by-second viewer retention rates and click-through rates in real time, making $F$ extremely high. In high-$F$ environments, senders can learn receivers' relevance structures and adaptively optimize their content.

ADI is defined as a function of these four variables:

\begin{equation}
ADI = \frac{G \cdot I \cdot F}{B}
\end{equation}

The three variables constituting the numerator ($G$, $I$, $F$) are all factors promoting agenda decentralization, while $B$ in the denominator is a factor suppressing it. The multiplicative combination of the four variables has theoretical significance. If any one variable is near zero, ADI as a whole is suppressed no matter how large the other variables are. This property becomes important in the cable-era analysis below.

For operational use, we adopt the convention that each of $G$, $I$, $F$ is normalized to the unit interval $[0,1]$ (relative to the historical maximum on each dimension), and $B$ is bounded below by a small positive floor $B \geq \varepsilon > 0$ to avoid the singularity as $B \to 0$. Under this convention ADI is a bounded, dimensionless index in $[0, 1/\varepsilon]$, permitting comparison across historical stages and empirical estimation once operational proxies for each variable are chosen (see Section~\ref{sec:discussion} for a discussion of empirical measurement).

\subsection{Trajectory of ADI Across Media History}

Using the ADI framework, the major stages of media history can be described as continuous change.

The broadcast monopoly era (1950s--1980s) was a state where ADI was near its minimum. Under broadcast licensing, $B$ was extremely high, and with channel numbers limited to a few stations, $G$ was extremely low. Broadcasting was unidirectional with $I$ near zero, and receivers' responses were aggregated only the next day as program-level ratings, making $F$ low as well. In this environment, a small number of media organizations effectively monopolized agenda-setting power, and the theory of \citet{mccombs1972agenda} was premised on precisely these conditions.

In the cable and satellite era (1980s--2000s), $B$ decreased and $G$ also rose somewhat with the increase in channel numbers. The post-broadcast democracy analyzed by \citet{prior2007post} corresponds to this stage, and the finding that increased media options enabled news avoidance can be understood as a consequence of rising $G$. But why did ADI not rise substantially during this period? The reason is that $I$ and $F$ remained low. Cable television was still a unidirectional broadcast medium, and receivers had no means to participate in agenda formation. Individual channels could obtain viewing data to some degree, but its granularity and speed were incomparable to YouTube analytics. Because ADI's four variables combine multiplicatively, even if $B$ and $G$ changed, as long as $I$ and $F$ remained near zero, the overall value remained suppressed. Agenda-setting power merely shifted from a few large organizations to many mid-sized organizations; the structure of power itself did not change.

A qualitative rupture occurred in the YouTube era (2005--present). The establishment of an advertising-revenue model realized a structure where neither senders nor receivers bear costs, and $B$ fell to effectively zero. Simultaneously, granularity refined to individual videos ($G$ surging), bidirectional participation through comments, sharing, and response videos ($I > 0$), and real-time second-by-second analytics ($F$ surging) all occurred together. The simultaneous change of all four variables in the same direction caused ADI to rise sharply, creating a qualitative rupture with the cable era. The ``long tail'' phenomenon discussed by \citet{anderson2006long}---the structure whereby not only a few hit products but also countless niche contents become economically viable---can be understood as an economic consequence of precisely this high-ADI environment. This analysis demonstrates the importance of grasping changes in the media environment as simultaneous changes in multiple structural variables rather than attributing them to a single technological factor such as ``the advent of the Internet.''

\subsection{Formalization of Bridging Mechanisms}
\label{sec:bridging-form}

The decentralization of agenda-setting power does not by itself guarantee the health of the social information environment. Rising ADI makes pluralistic information dissemination possible while simultaneously potentially creating states where different information spheres are mutually disconnected. Realizing a healthy information environment requires bridging mechanisms through which information moves between different information spheres to be functioning alongside decentralization.

Bridging mechanisms can be classified into three types based on their mode of occurrence. The first type is serendipitous bridging ($C_s$), referring to opportunities for individuals to encounter information outside their own information sphere unintentionally. Random encounters in urban public spaces, conversations with colleagues from different departments at work, and televisions playing in waiting rooms exemplify this type. Passive viewing of mass media was also a form of serendipitous bridging, and the ``chance encounters'' that \citet{sunstein2001republic} argued were essential for democracy essentially addressed this type. Serendipitous bridging requires no institutional design but depends on sharing physical space, and it has been declining with the spread of remote work and personalization.

The second type is institutional bridging ($C_{inst}$), referring to institutional arrangements that structurally bring together people of different positions. Parliamentary debates, the principle of public trials, jury and lay judge systems, and deliberative polling exemplify this type. These institutions are intentionally designed to facilitate information exchange among individuals belonging to different echo chambers within institutional frameworks, distinguished from the first type by not depending on serendipity. However, like serendipitous bridging, they share the common feature of presupposing physical co-presence, constraining their scalability.

The third type is algorithmic bridging ($C_a$), referring to arrangements that expose users to information outside their own information sphere through the design of recommendation algorithms. Bridging-based ranking by \citet{ovadya2023bridging} and Community Notes on Twitter (now X) can be positioned as concrete implementations of this type. Algorithmic bridging has qualitatively different properties from the previous two. It does not require physical co-presence and can be deployed at scale, and it can leverage feedback resolution $F$ to adaptively improve bridging methods.

Here it is necessary to make explicit the ambivalent role that $F$ plays vis-\`{a}-vis algorithmic bridging. As a component of ADI, $F$ enables senders to optimize for receiver engagement, consequently promoting the deepening of echo chambers. At the same time, however, because receiver reactions to bridging information are also measurable in real time, $F$ also serves as a data foundation for learning which types of bridging information are more readily accepted by which users and adaptively optimizing bridging design. $F$ is simultaneously the cause of echo chambers and the technical condition for their prescription. To formalize this ambivalence, we write algorithmic bridging as $C_a = h(\lambda_k, F)$, where $\lambda_k$ is the bridging weight intentionally set by the algorithm designer for information domain $k$ and $F$ is the data foundation for implementing and measuring the effectiveness of that design. If $\lambda_k = 0$, no bridging occurs no matter how high $F$ is; if $F = 0$, effective implementation is impossible no matter what $\lambda_k$ is set.

\subsection{The Social Information Health (SIH) Model}

Integrating the above discussion, we define Social Information Health (SIH) by combining the degree of agenda-setting power decentralization (ADI) with the strength of bridging mechanisms ($C$):

\begin{equation}
SIH = ADI \times C = \frac{G \cdot I \cdot F}{B} \times \left(\alpha C_s + \beta C_{inst} + \gamma C_a\right)
\end{equation}

\noindent where $C_a = h(\lambda_k, F)$ as defined in Section~\ref{sec:bridging-form}. $\alpha$, $\beta$, $\gamma$ are coefficients representing the relative weight of each bridging type. This formulation has two important implications. One is that if ADI is high but $C$ is low, SIH is low---a decentralized but disconnected information environment. The other is that even if ADI is low, SIH can be maintained if $C$ is at a certain level---corresponding to the situation where serendipitous bridging played a role during the mass media era.

The challenge currently facing the information environment can be clearly described from the perspective of this model. While ADI has reached its highest level in history, $C_s$ is declining due to remote work and personalization, $C_{inst}$ is becoming dysfunctional with declining institutional trust, and $C_a$ has not yet adequately developed. The development of $C$ has not kept pace with the rise of ADI. The paradox frequently noted---that ``information has been democratized yet division deepens''---is structurally explained as this asynchronous change between ADI and $C$.

\subsection{Optimal Allocation of Bridging Resources}

What is critically important in the design of bridging mechanisms is the recognition that the need for bridging is not uniform across all information domains. Using the two variables introduced in Section~\ref{sec:evolutionary}---verifiability $V_k$ and collective scope $S_k$---we can derive the optimal value of the algorithmic bridging weight $\lambda_k$ for each information domain $k$:

\begin{equation}
\lambda_k^* = \lambda_0 \cdot \min\!\left(\frac{S_k}{V_k},\; \rho_{\mathrm{sat}}\right)
\end{equation}

\noindent where $\lambda_0 > 0$ is a global scale and $\rho_{\mathrm{sat}}$ is a domain-independent saturation cap calibrated so that the resulting $\lambda_k^*$ does not exceed the saturation point $\lambda^*$ derived from Section~4.6 (empirically estimated at $\lambda^* \approx 0.21$ in Section~\ref{sec:simulation}). The formula thus has two components: an $S/V$-based ranking of domains, and an absolute-level ceiling that prevents wasted investment beyond the saturation regime. Without the ceiling, the earlier proportional form $\lambda_k^* \propto S_k / V_k$ would prescribe values far exceeding $\lambda^*$ in high-$S/V$ domains, which by the saturation model of Section~4.6 would produce no additional gain in bridging effect while unnecessarily reducing user satisfaction.

In information domains where $V$ is high and $S$ is low, such as hobbies and local lifestyle information, $\lambda_k^*$ takes values near zero. Users can verify information through their own direct experience, and the scope of decisions is limited to the individual, so even if echo chambers form in this domain, the social cost is small. Algorithmic intervention is unnecessary and would only needlessly sacrifice user engagement. In contrast, in information domains where $V$ is low and $S$ is high---national policy, security, public health, and the like---$\lambda_k^*$ takes large values, and active algorithmic bridging intervention is justified. In these domains, verification through individual direct experience is difficult while the quality of information-based collective decision-making affects society as a whole. If cognitive closure is left unaddressed, the quality of democratic decision-making itself may decline. In intermediate domains such as local government and regional economics, $\lambda_k^*$ takes moderate values; some bridging is beneficial, but intervention equivalent to the policy domain would be excessive.

This domain-specific optimization is a substantive extension of the bridging systems of \citet{ovadya2023bridging}. Their framework, while explicitly presenting many issues as open problems, does not systematically develop domain-differentiated bridging weights; the present model provides such a differentiation based on $S/V$. By introducing this differentiation, a highly feasible design guideline is obtained that achieves effective bridging in domains related to collective decision-making without undermining the user experience in hobby and lifestyle information.

\subsection{Saturation of Bridging Effects}

Increasing the bridging weight $\lambda_k$ without limit does not produce unlimited increases in effect. Based on the evolutionary cognitive constraints discussed in Section~\ref{sec:evolutionary}, information with low relevance is not processed but simply ignored. Even if the supply of bridging information is increased, the amount that receivers can actually process and integrate is bounded by cognitive resource constraints. Supply beyond this bound does not cause harm but is simply ignored.

We formalize this saturation property as follows:

\begin{equation}
E(C_k) = C_{\max,k} \cdot \left(1 - e^{-\kappa_k \cdot C_k}\right)
\end{equation}

$C_{\max,k}$ is the upper limit of bridging effect in information domain $k$, and $\kappa_k$ is a parameter governing the rate of saturation. This asymptotic saturation model has clear policy implications regarding investment in bridging. If bridging is insufficient, cognitive closure is left unaddressed, producing the substantive problem of declining collective decision-making quality. However, additional investment beyond a certain level produces almost no effect; it is not harmful but merely inefficient. As a consequence of adopting an asymptotic saturation model rather than an inverted-U model (too much bridging is harmful), the practical guideline derived for bridging design is: ``there is no need to fear excessive intervention, but investment beyond the optimal level wastes resources.''

The next section verifies the dynamic behavior of the model constructed so far through agent-based simulation. By comparing three algorithm designs---no bridging, uniform bridging, and domain-selective bridging---we quantitatively demonstrate that domain-selective bridging most efficiently improves SIH.

\section{Agent-Based Simulation: Dynamic Analysis and Model Comparison}
\label{sec:simulation}

\subsection{Simulation Design}

To verify how the ADI-SIH model constructed in the previous section behaves dynamically under different algorithm designs, we conducted an agent-based model (ABM) simulation. The purpose of this simulation is to compare three algorithm designs---no bridging, uniform bridging, and domain-selective bridging---under identical conditions, and to quantitatively demonstrate the superiority of domain-selective bridging in simultaneously achieving social information health and user satisfaction. The simulation code was implemented in Python, and is publicly available on an OSF repository for reproducibility.

The simulation is designed to test four hypotheses derived from the theoretical model:

\begin{description}
\item[H1 (Model~0):] Under pure engagement optimization, echo chamber depth (ECD) rises monotonically and cross-cluster information sharing (CIS) declines, while user satisfaction (US) is maintained at a high level.
\item[H2 (Model~1):] Uniform bridging raises CIS across all domains but sharply reduces US, particularly in low-$S/V$ domains where bridging is not warranted.
\item[H3 (Model~2):] Domain-selective bridging achieves CIS gains comparable to or exceeding Model~1 in high-$S/V$ domains, while maintaining US at Model~0 levels in low-$S/V$ domains.
\item[H4 (Saturation and $F$):] The effect of $\lambda$ on CIS exhibits asymptotic saturation, and feedback resolution $F$ acts as a substrate that either accelerates echo chamber deepening (when $\lambda = 0$) or improves bridging precision (when $\lambda > 0$).
\end{description}

The model comprises $N = 300$ agents, each possessing an interest vector $\mathbf{p}_j \in \mathbb{R}^2$ located in a two-dimensional interest space. As the initial configuration, four clusters were set up with cluster centers placed at equal intervals on a circle of radius 4.0. Agents within each cluster were dispersed according to a normal distribution with standard deviation 0.8 from the center. This configuration represents an initial state in which groups belonging to different information spheres are clearly separated.

At each time step, 30 information items are generated. Sixty percent of items are generated near randomly selected agents (corresponding to information circulating within existing echo chambers), 20\% at intermediate positions between agents belonging to different clusters (corresponding to bridging information), and the remaining 20\% are placed at random positions in interest space. This composition reflects the fact that the information environment consists of a mixture of in-echo-chamber information, bridging information, and noise. For each agent, the top 5 items are recommended based on the scoring function described below.

For opinion dynamics, we adopted the bounded confidence model \citep{deffuant2000mixing,hegselmann2002opinion}. After consuming a recommended item, an agent updates its interest vector at an update rate of $\mu = 0.02$ toward the item if it falls within the confidence threshold ($\delta = 2.0$) of the agent's interest vector. For items exceeding the threshold, the update rate is attenuated to one-tenth. This design reflects the cognitive constraint that humans do not readily accept information that deviates extremely from their existing belief systems, incorporating the realistic condition that echo chambers are not immediately dissolved simply because bridging information is recommended. The simulation was run for 150 time steps.

\subsection{Three Algorithm Designs and Measurement Metrics}

The three algorithm designs compared are all formalized as differences in the scoring function for recommendations.

Model~0 (no bridging) corresponds to current engagement-maximizing algorithms, defining the recommendation score as $\text{Score}(i, j) = 1 / (1 + \|\mathbf{p}_j - \mathbf{q}_i\|^2)$. Items closer to the agent's interest vector receive higher scores in this pure engagement optimization, with bridging weight $\lambda = 0$. Model~1 (uniform bridging) formalizes the prescriptions of \citet{sunstein2001republic} and \citet{ovadya2023bridging}. It adds a bridging term to the engagement term in the recommendation score, assigning a bridging score proportional to distance for items beyond a threshold $\theta = 1.5$ from the agent's interest vector. The bridging weight is set uniformly at $\lambda = 0.4$ across all information domains. Model~2 (domain-selective bridging) is the proposal of this paper, differentiating the bridging weight by information domain characteristics using $\lambda_k = \lambda_0 \cdot S_k / V_k$ ($\lambda_0 = 0.5$). From this formulation, the hobby domain ($V = 0.9, S = 0.1$) has $\lambda = 0.056$, the local government domain ($V = 0.5, S = 0.5$) has $\lambda = 0.50$, and the national policy domain ($V = 0.1, S = 0.9$) has $\lambda = 4.50$. These $(V, S)$ values are chosen as representative illustrative endpoints for three qualitatively distinct domain types; they are not empirical measurements, and Section~\ref{sec:discussion} discusses how they should be estimated empirically in future work. Note that $\lambda = 0.50$ and $\lambda = 4.50$ substantially exceed the saturation point $\lambda^* \approx 0.21$ identified in Section~\ref{sec:simulation}. As theoretically predicted by the saturation model, the CIS values produced in these domains under Model~2 correspond to the plateau of the saturation curve; the clipped optimum $\lambda_k^* = \lambda_0 \cdot \min(S/V, \rho_{\mathrm{sat}})$ introduced in Section~4.5 would yield the same CIS plateau at lower US cost, and we treat Model~2's simulation results as an upper bound on the CIS component and a lower bound on the US component for these domains. The reported qualitative dominance ranking (Model~2 $\succ$ Model~0, Model~1) is therefore robust to this refinement.

The simulation was executed independently for each of the three models across the above three information domains (9 conditions total), and results were evaluated using the following metrics. Echo Chamber Depth (ECD) is calculated as the ratio of between-cluster variance to total variance of agents' interest vectors; values closer to 1 indicate greater convergence of agents within each cluster, signifying deepened echo chambers. Cross-Cluster Information Sharing (CIS) is defined as the inverse of the average distance between cluster centers, $CIS = 1/(1 + \bar{d})$, capturing the degree to which agents in different clusters share similar information. User Satisfaction (US) is defined as the mean engagement score for recommended items, reflecting the trade-off in bridging design as it tends to decrease with increased bridging. In addition to these three metrics, SIH was calculated as $SIH = \sum_k (S_k / V_k) \cdot CIS_k$ to evaluate the overall quality of the information environment across the three models.

\subsection{Main Results}

The simulation results supported all four hypotheses. The findings corresponding to each hypothesis are described below.

\begin{figure}[H]
\centering
\includegraphics[width=\textwidth]{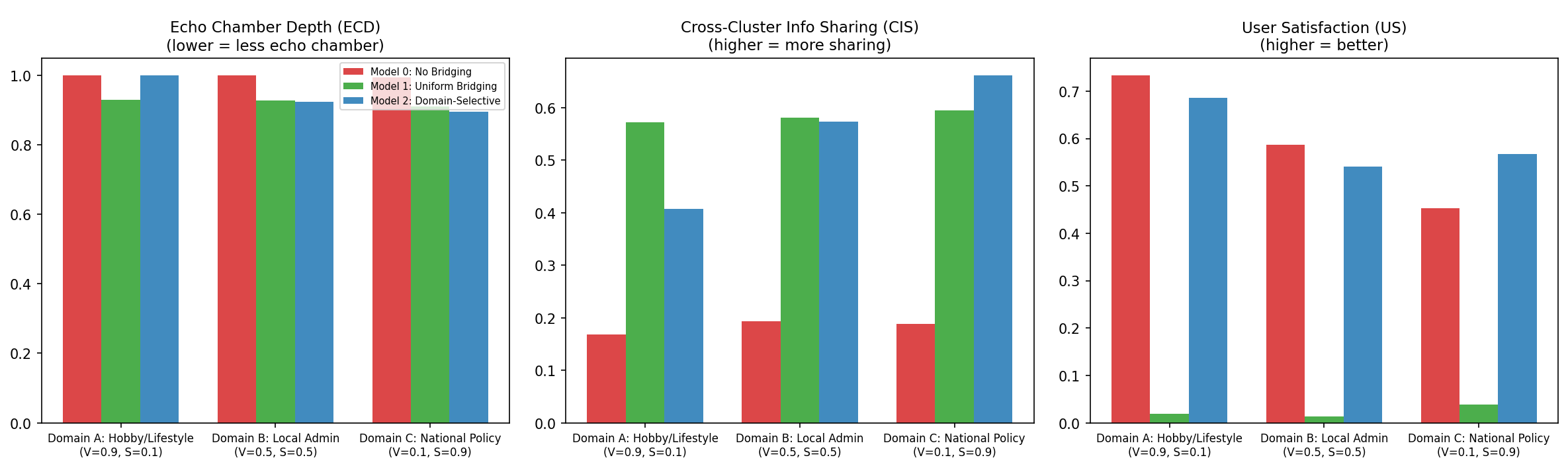}
\caption{Comparison of three algorithm designs across information domains. Left: Echo Chamber Depth (ECD). Center: Cross-Cluster Information Sharing (CIS). Right: User Satisfaction (US).}
\label{fig:comparison}
\end{figure}

In Model~0 (no bridging), ECD reached approximately 1.0 across all domains, with agents converging completely within their initial clusters. CIS remained at 0.17--0.19 across all domains, with almost no information sharing between different clusters. On the other hand, US maintained the highest level among all models (full-domain average of 0.59). This result shows that algorithms maximizing only engagement secure short-term user satisfaction while leaving cognitive closure unaddressed in all domains, including those necessary for collective decision-making.

Model~1 (uniform bridging) substantially improved CIS across all domains to 0.57--0.59, successfully promoting inter-cluster information sharing. However, this improvement was accompanied by a significant cost. US plummeted across all domains, with the average falling to 0.023. Because bridging was applied uniformly even in hobby domains where its need is low, users were recommended large quantities of information divergent from their interests, severely damaging the user experience. This finding quantitatively demonstrates that \citet{sunstein2001republic}'s ``chance encounter'' prescription and \citet{ovadya2023bridging}'s uniform bridging-based ranking, while theoretically justified, carry the risk of causing user attrition that eliminates bridging opportunities altogether.

In contrast, Model~2 (domain-selective bridging) achieved both CIS improvement and US maintenance. In the national policy domain (Domain~C, $\lambda = 4.50$), CIS reached 0.66, the highest value among all three models, exceeding Model~1's information sharing. Simultaneously, in the hobby domain (Domain~A, $\lambda = 0.056$), US was maintained at 0.69, and the full-domain average US was 0.60, nearly equivalent to Model~0 (0.59). By differentiating bridging weights according to information domain characteristics, a design was realized that actively bridges in domains related to collective decision-making while not undermining the user experience in hobby domains.

\begin{figure}[H]
\centering
\includegraphics[width=0.7\textwidth]{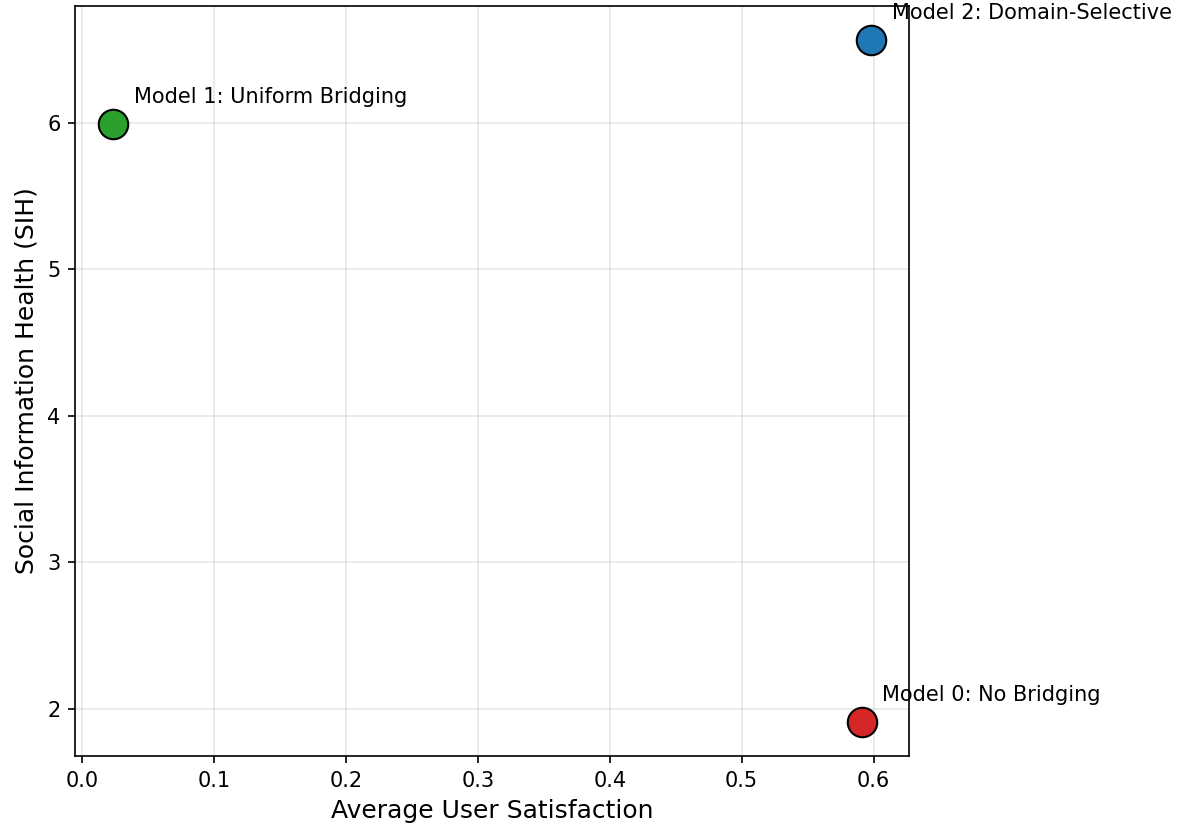}
\caption{Trade-off between user satisfaction and social information health. Model~2 (domain-selective) dominates the upper-right quadrant.}
\label{fig:tradeoff}
\end{figure}

A comprehensive comparison of the three models can be visualized by placing each model in the two-dimensional space of SIH (social information health) and US (user satisfaction) (Figure~\ref{fig:tradeoff}). Model~0 is positioned in the lower right (high US, low SIH), Model~1 in the upper left (high SIH, near-zero US), and Model~2 in the upper right (highest SIH with US equivalent to Model~0). The efficiency index, defined as the product of SIH and US, was 1.13 for Model~0, 0.14 for Model~1, and 3.93 for Model~2. Model~2's efficiency is approximately 3.5 times that of Model~0 and approximately 28 times that of Model~1. We report these ratios with an explicit caveat: because Model~1's US collapses to 0.023, the denominator of the Model~2/Model~1 ratio is small, and modest changes in Model~1's simulation parameters (e.g., the uniform $\lambda$ from 0.4 to 0.2) shift the ratio substantially without changing the qualitative dominance ranking. Our primary claim is directional: Model~2 strictly dominates both Model~0 (higher SIH at equivalent US) and Model~1 (higher US at equivalent or higher SIH). The raw pairs (SIH, US) rather than their product are the more informative comparison; the 28$\times$ figure should be read as an illustrative summary of the joint dominance, not as a precise structural constant.

\subsection{Dynamic Analysis}

\begin{figure}[H]
\centering
\includegraphics[width=\textwidth]{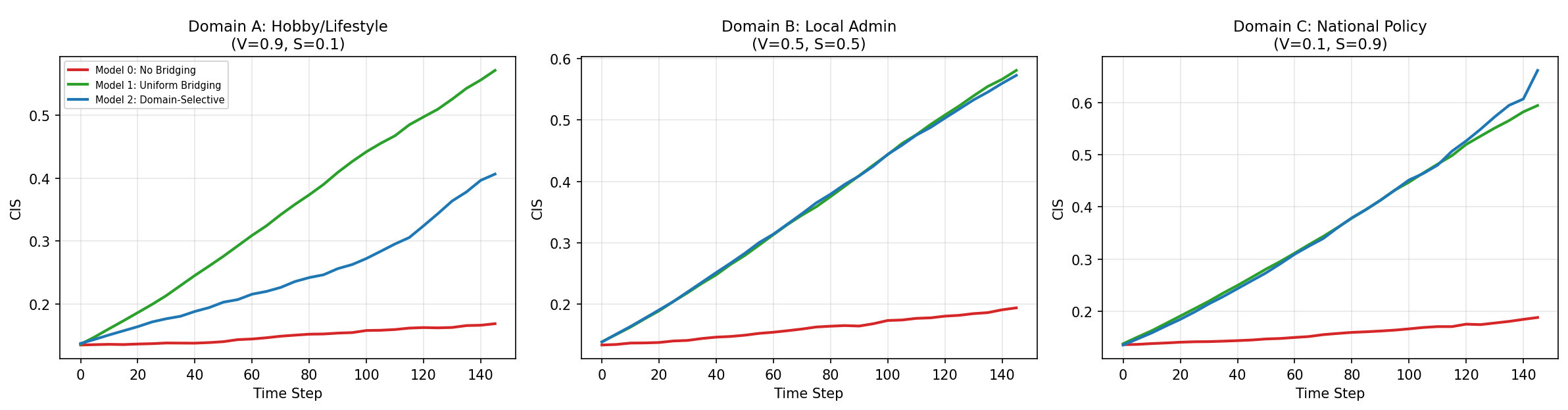}
\caption{CIS dynamics over time by information domain.}
\label{fig:dynamics}
\end{figure}

Tracking the temporal evolution of CIS reveals the qualitative differences in dynamic behavior among the three models.

In Model~0, CIS continued to decline monotonically over time. This behavior is explained by the positive feedback loop generated by engagement optimization. Agents' exposure only to information within their own cluster causes their interest vectors to converge toward cluster centers, which in turn causes only within-cluster information to receive high scores for recommendation at the next time step. This feedback loop does not self-terminate, and without external intervention, CIS continues to decline without reaching a steady state. This result suggests that echo chamber deepening is not a side effect of algorithm design but a logical necessity of engagement optimization.

In Models~1 and 2, CIS showed a pattern of rapid initial increase followed by stabilization. In particular, in Model~2's Domain~C, the high bridging weight of $\lambda_C = 4.50$ resulted in the most rapid stabilization at the highest level (CIS = 0.66). Notably, in Model~2's Domain~A (hobbies), CIS rose more slowly than in Model~1. This is a natural consequence of the small bridging weight $\lambda_A = 0.056$, but the fact that the insufficiency of bridging in Domain~A does not produce social costs is theoretically predicted by the low $S/V$ (= 0.11), and the fact that CIS remains low in this domain is itself not a problem.

\subsection{Saturation of Bridging Effects}

\begin{figure}[H]
\centering
\includegraphics[width=0.7\textwidth]{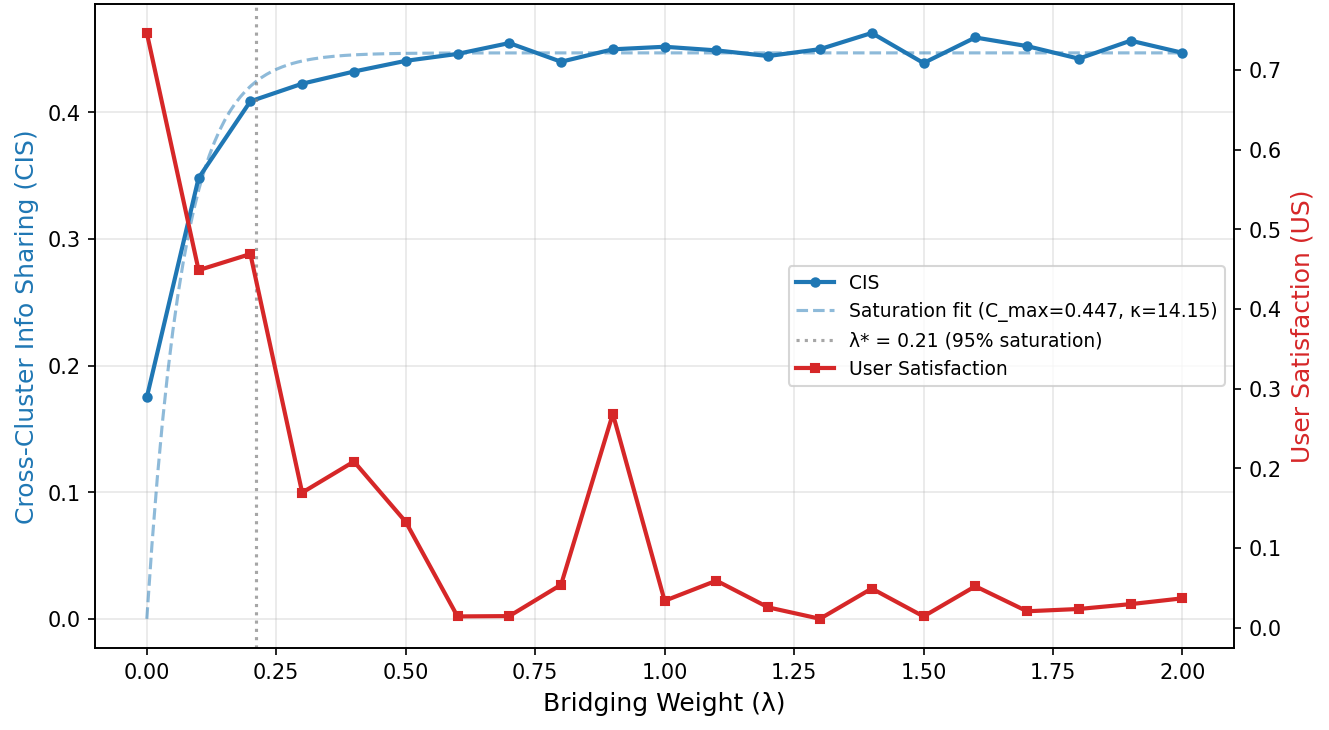}
\caption{Effect of bridging weight ($\lambda$) on CIS and user satisfaction in Domain~C (national policy). CIS saturates at $\lambda^* = 0.21$.}
\label{fig:saturation}
\end{figure}

To verify the theoretically predicted asymptotic saturation of bridging effects from Section~\ref{sec:model}, we analyzed changes in CIS and US when $\lambda$ was varied from 0 to 2.0 in Domain~C.

CIS rose rapidly with increasing $\lambda$, entering the saturation region around $\lambda \approx 0.2$. Fitting the saturation curve to $E(C) = C_{\max}(1 - e^{-\kappa C})$ yielded $C_{\max} = 0.447$ and $\kappa = 14.15$, with the $\lambda$ value at which CIS reaches 95\% of its theoretical maximum estimated at $\lambda^* = 0.21$. Increases in $\lambda$ beyond this saturation point contributed almost nothing to CIS improvement. Increasing $\lambda$ approximately tenfold from 0.21 to 2.0 produced only a few percentage points of CIS improvement.

Meanwhile, US declined rapidly with increasing $\lambda$. From 0.75 at $\lambda = 0$, US fell to 0.17 around $\lambda = 0.3$, and hovered near zero for $\lambda > 0.5$. Superimposing the CIS saturation curve and the US decline curve confirms visually that a region near $\lambda^*$ exists where CIS improvement and US maintenance are best reconciled.

This result is consistent with the theoretical prediction that bridging effects saturate asymptotically. As a policy implication, underinvestment in bridging leaves cognitive closure unaddressed, but overinvestment far beyond $\lambda^*$ brings no additional CIS improvement and only reduces US. While the Section~\ref{sec:model} prediction that underinvestment is problematic and overinvestment is not harmful but merely inefficient holds, factoring in the cost of US decline reveals that overinvestment carries practical disadvantages beyond inefficiency, including the risk of user attrition.

\subsection{Effects of Feedback Resolution}
\label{sec:F-effects}

\begin{figure}[H]
\centering
\includegraphics[width=\textwidth]{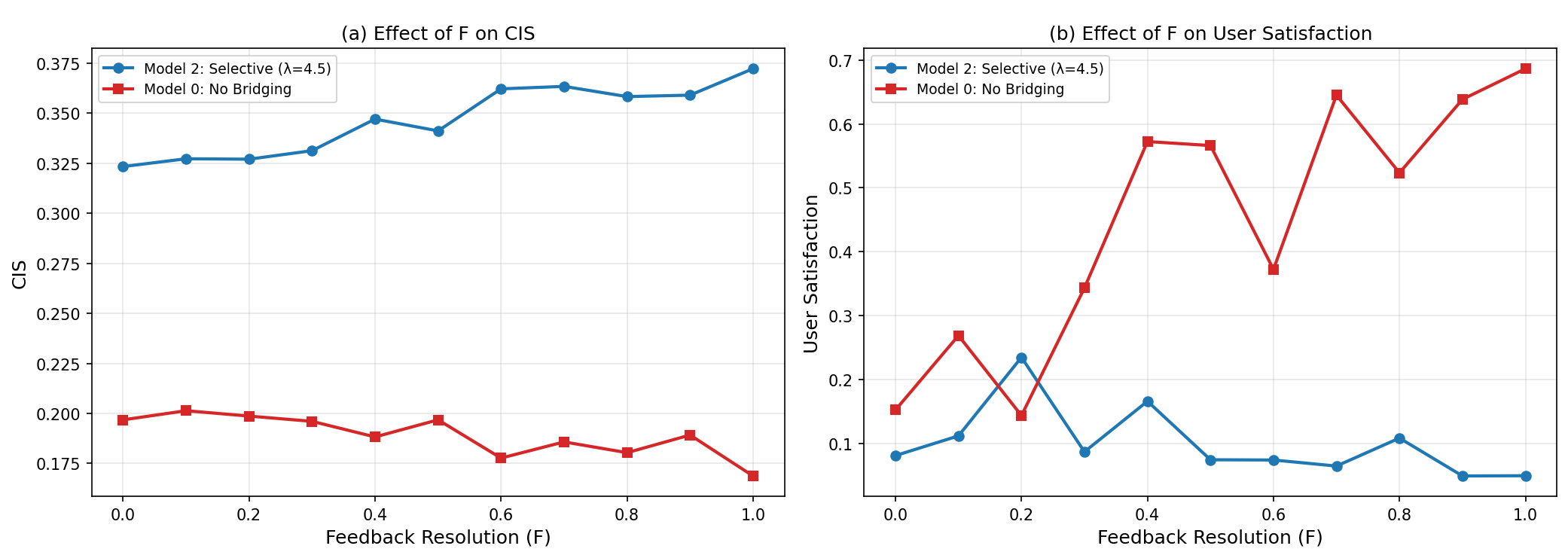}
\caption{Effect of feedback resolution ($F$) on CIS and user satisfaction. In Model~0, higher $F$ deepens echo chambers; in Model~2, higher $F$ improves bridging precision.}
\label{fig:feedback}
\end{figure}

In Section~\ref{sec:model}, algorithmic bridging was formalized as $C_a = h(\lambda_k, F)$, and it was argued that feedback resolution $F$ functions as the data foundation for implementing and improving bridging. To verify this theoretical prediction, we conducted a simulation explicitly manipulating $F$. $F$ was modeled as the precision with which the algorithm observes agents' interest vectors. At $F = 1.0$, the algorithm grasps agents' exact interest positions (corresponding to perfect analytics); at $F = 0.0$, the algorithm has almost no information about agents' interest positions (corresponding to coarse aggregate feedback). The lower $F$ is, the more noise contaminates the estimation of interest vectors underlying recommendations.

The results clearly corroborated the theoretical prediction regarding the ambivalence of $F$. In Model~0 (no bridging), CIS declined slightly with increasing $F$ (CIS = 0.20 at $F = 0.0$, CIS = 0.17 at $F = 1.0$). Higher $F$ improved the precision of engagement optimization, causing within-cluster information to be more accurately recommended to each agent, thereby deepening echo chambers more efficiently. Simultaneously, US rose dramatically ($F = 0.0$: 0.15; $F = 1.0$: 0.69). This succinctly demonstrates why current platforms have incentives to pursue high $F$. Improving $F$ dramatically improves user satisfaction, but without bridging design, that improved precision serves only to deepen echo chambers.

In contrast, in Model~2 (domain-selective bridging), CIS improved with increasing $F$ ($F = 0.0$: CIS = 0.32; $F = 1.0$: CIS = 0.37). With higher $F$, the algorithm can more precisely identify each agent's interest position and select bridging information at appropriate distances that do not exceed the confidence threshold. The formulation $C_a = h(\lambda_k, F)$---that bridging design intent ($\lambda_k > 0$) is effectively implemented by the data foundation ($F$)---was empirically supported.

This result also has important policy implications. Restricting $F$ has the effect of suppressing echo chamber deepening while simultaneously reducing bridging precision, creating a trade-off. Rather than targeting $F$ for regulation, extending the scope of $F$'s application to include bridging design would be a more rational regulatory direction for improving the information environment.

\subsection{Visualization of Agent Distributions}

\begin{figure}[H]
\centering
\includegraphics[width=\textwidth]{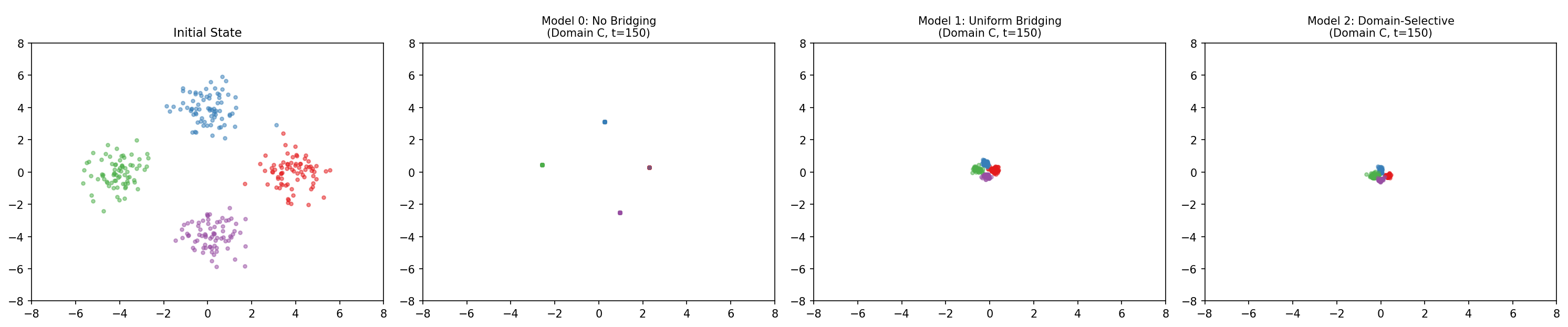}
\caption{Agent distribution in interest space: initial state and final state for each model (Domain~C).}
\label{fig:scatter}
\end{figure}

Finally, visualizing changes in agent distributions in interest space intuitively shows the qualitative differences in behavior among the three models. In the initial state, four clusters are clearly separated, but distributions after 150 time steps differ substantially by model. In Model~0, cluster separation is maintained while agents within each cluster condense toward centers, visually confirming echo chamber deepening. In Model~1, diffusion occurs uniformly across all domains, with cluster boundaries becoming indistinct; this means that users' interest structures are unnecessarily disrupted even in hobby domains. In Model~2, the distance between clusters is notably reduced in Domain~C, with agent mixing progressing, while the initial cluster structure is largely maintained in Domain~A. This contrast is visual evidence that domain-selective bridging distinguishes between domains requiring bridging and those that do not.

\section{Discussion}
\label{sec:discussion}

\subsection{Theoretical Implications}

The analysis in this paper proposes a fundamental shift in perspective for the existing debate on echo chambers.

In prior discussions, echo chambers have been positioned as pathologies produced by algorithm design or user behavioral choices. \citet{sunstein2001republic,sunstein2017republic} problematized the loss of opportunities for citizens to experience chance encounters, and \citet{pariser2011filter} criticized the filter bubbles created by personalization algorithms. Common to these arguments is a structure that implicitly sets the mass media era, when all citizens shared a single information environment, as the baseline and treats deviation from it as a problem.

This paper has questioned this baseline itself. The human cognitive system is designed to preferentially process information of high relevance at the small-group scale in evolutionary environments. As relevance theory \citep{sperber1995relevance} shows, since information processing priorities are determined by the ratio of cognitive effects to processing costs, assigning low priority to spatially and socially distant information is normal operation of the cognitive system, not a malfunction. National mass media's ability to impose a single agenda on the entire citizenry was an exceptional state possible only under the historically specific condition of information channel scarcity. Echo chamber formation should be understood as the re-emergence of human cognitive constraints after this exceptional state was resolved.

This repositioning enables treating echo chambers not as ``a problem to be solved'' but as ``a design precondition.'' The focus of the problem shifts from the existence of echo chambers themselves to the insufficiency of bridging mechanisms in information domains where verifiability is low and collective scope is large. This shift fundamentally changes the direction of prescriptions. The enlightenment-style prescription of ``expose yourself to diverse information'' has low feasibility because it demands something that contradicts human cognitive constraints, but the prescription of incorporating bridging into algorithmic scoring functions adjusts the information environment while accepting cognitive constraints as given, and therefore has higher feasibility.

The ADI-SIH model provides a quantitative framework for this argument. $ADI = G \cdot I \cdot F / B$ enables continuous description of the degree of agenda-setting power decentralization, and each stage of media history can be unified from the cable era's ``partial decentralization'' to the YouTube era's ``complete decentralization.'' In particular, the ability to explain why ADI did not rise substantially during the cable era---through multiplicative suppression due to low $I$ and $F$---provides a structural answer to the question ``why did no qualitative transformation occur despite multi-channel proliferation before the Internet?''

\subsection{Contributions to the Bridging Systems Discussion}

This paper has extended the bridging systems discussion of \citet{ovadya2023bridging} in three directions.

First, it provided a theoretical foundation for differentiating the need for bridging by information domain. The framework that derives bridging weights as $\lambda_k^* \propto S_k / V_k$ using the two variables of verifiability $V$ and collective scope $S$ provides a principled answer to the question ``where should bridging be deployed?'' as a resource allocation problem. The simulation results demonstrated that this domain-selective bridging is overwhelmingly superior to uniform bridging in simultaneously achieving social information health and user satisfaction.

Second, it theoretically predicted the asymptotic saturation of bridging effects and confirmed this through simulation. The finding that 95\% saturation is reached at $\lambda^* = 0.21$ suggests that a ``sufficient level'' exists for investment in bridging, carrying the practically important implication that unlimited increases in bridging are unnecessary.

Third, by grounding echo chamber formation in evolutionary cognitive constraints, it made explicit the principled constraint that bridging design must be compatible with cognitive constraints to function. The finding by \citet{bail2018exposure} that exposure to opposing political views can strengthen polarization, particularly among Republicans, is naturally understood within the bounded confidence model framework. Exposure to information exceeding the confidence threshold produces almost no update to interest vectors and may even provoke backlash. This simulation suggests that bridging design requires adjustment not only of information ``quantity'' but also of ``distance'' that does not exceed receivers' confidence thresholds.

\subsection{Policy Implications}

The findings of this paper provide several specific suggestions for policy discussions on algorithm regulation.

First is the suggestion that the regulatory targets for echo chambers should be differentiated by information domain. This suggestion also connects to the ``news desert'' problem created by the decline of local newspapers \citep{abernathy2018expanding}. The disappearance of local papers means that bridging (particularly serendipitous bridging $C_s$ and institutional bridging $C_{inst}$) is lost in the local information domain where $V$ is high and $S$ is moderate, suggesting that domains exist where algorithmic bridging alone cannot serve as a substitute. Regulatory intervention in hobby and lifestyle echo chambers is unnecessary, and regulatory resources should be concentrated on information domains related to collective decision-making. This distinction can be principally derived through the $S/V$ ratio, providing an objective criterion for regulatory authorities to identify target domains.

Second, as an algorithm design guideline for platform operators, we can propose incorporating a Bridging term into scoring functions and its domain-specific optimization. As the simulation showed, uniform bridging carries the risk of collapsing user satisfaction, giving platform operators little incentive to voluntarily adopt it. Domain-selective bridging improves social information health while maintaining user satisfaction, potentially serving as a design guideline acceptable to operators as well.

Third, there are policy implications regarding feedback resolution $F$. As the simulation (Section~\ref{sec:F-effects}) showed, rising $F$ produces diametrically opposite consequences depending on whether bridging design is present. In environments without bridging design, improved $F$ accelerates echo chamber deepening, while in environments with bridging design, it improves bridging precision. Therefore, a regulatory approach that restricts $F$ creates the trade-off of also hindering bridging improvement. A more rational regulation would be mandating that $F$ be applied to bridging design as well, rather than restricting $F$ itself.

\subsection{Reinterpreting the Decline in User Satisfaction: Removal of the Entertainment Frame}

The simulation results (Section~\ref{sec:F-effects}) showed that when domain-selective bridging is introduced, user satisfaction (US) declines in the national policy domain (Domain~C) while remaining high in the hobby domain (Domain~A), producing a bifurcated user experience across information domains. This divergence is amplified on platforms with higher feedback resolution $F$. As engagement optimization precision increases, satisfaction in hobby domains improves, while in the policy domain where bridging has been introduced, ``information that does not match one's preferences but is important'' is recommended.

At first glance, this result suggests a risk that users will selectively disengage from the policy domain. However, evaluating this risk requires considering what current engagement-optimization algorithms are doing to policy information.

Current engagement-only algorithms process and deliver policy information so that it is consumed as entertainment. Emotionally provocative headlines, theatrical staging of conflict, and steering users toward commentary from their preferred political perspective---these techniques maintain engagement. This constitutes an operation of disguising information that should properly be consumed for factual acquisition within a frame of entertainment consumption; we term this the ``entertainment frame.'' Consumption of policy information through the entertainment frame is the primary mechanism that deepens echo chambers. As long as consumption is possible as ``pleasurable confirmation of one's political position,'' bridging information is excluded and cognitive closure progresses. Policy content with high US is, in essence, unbridged content.

The introduction of domain-selective bridging renders this entertainment frame dysfunctional. The insertion of bridging information makes it difficult to consume policy information as ``entertainment for confirming one's existing beliefs.'' The expected consequence is a change in users' mode of engagement with policy information---a shift from entertainment consumption to factual information acquisition.

The contrast with disaster information clarifies this point. Disaster information is a paradigmatic example of information that users actively seek out despite its low entertainment value, driven by practical necessity. Information about national policy, security, and epidemiological facts should similarly be consumed in this factual-acquisition mode, but current engagement optimization has been converting it into entertainment mode. Domain-selective bridging, by making this conversion impossible, may have the effect of facilitating a return to the factual-acquisition mode.

Accordingly, the decline in US observed in the simulation should be reinterpreted not as a ``cost of bridging'' but as ``removal of the entertainment frame.'' Policy content with low engagement has low entertainment value but high informational value. The US metric measures engagement but does not measure the democratic function of information. This distinction carries important implications for the design of algorithm evaluation metrics as well. CIS (cross-cluster information sharing) in the SIH model captures this democratic function to some degree but does not reflect changes in users' information consumption modes. Developing metrics that capture changes in information consumption modes remains a task for future research.

We emphasize that the entertainment-frame argument is a theoretical proposal that goes beyond the scope of the present simulation: the ABM does not model users' consumption \emph{mode} (entertainment vs.\ factual acquisition), only their engagement response. The claim that domain-selective bridging shifts consumption modes is therefore an empirically testable hypothesis flagged for future work rather than a result of the current simulation.

This discussion also has important implications for the media supply side. Maintaining the entertainment frame has imposed increasing production costs and ethical burdens on media organizations. An environment in which resources can be devoted to accurately conveying facts rather than to techniques for crafting emotionally provocative headlines or staging conflict represents a return to the fundamental function of journalism. Domain-selective bridging, through transforming users' information consumption modes, may also promote a transformation in the supply side's reporting modes.

\subsection{Limitations and Future Directions}

The analysis in this paper has several important limitations.

First, this paper is based on a theoretical model and simulation and does not conduct empirical verification using actual platform user behavior data. Simulation parameters (number of clusters, confidence threshold, update rate, etc.) were set based on theoretical considerations, but the values of these parameters in actual information environments need to be estimated empirically. In particular, concrete measurement methods for $V$ and $S$ have not been established, and how to estimate these values for each information domain remains an empirical task. One approach would be to measure in surveys ``can you confirm the accuracy of this topic yourself?'' (subjective evaluation of $V$) and ``does your information affect society-wide decision-making on this topic?'' (subjective evaluation of $S$), but the correspondence between subjective evaluations and objective characteristics also requires examination.

Second, the simulation model abstracts away several important characteristics of information environments. In actual information environments, agents are active across multiple information domains, and bridging experiences in one domain can influence behavior in others. Information quality and reliability differences, strategic behavior of senders (intentional generation of bridging content or gaming), and movement between platforms are also not considered in this model. Building extended models incorporating these factors is a task for future research.

Third, the optimal allocation formulation $\lambda_k^* \propto S_k / V_k$ implicitly assumes that $V$ and $S$ are independent, but in practice these variables may be correlated. For example, policy information with large collective scope tends to have low verifiability due to its abstractness. Developing more refined optimal allocation models that account for this correlation structure is also a theoretical task for the future.

Fourth, this paper focused on the ``quantity'' of bridging but did not sufficiently discuss the ``quality'' of bridging. As the bounded confidence model suggests, bridging information that greatly exceeds the confidence threshold can actually strengthen polarization. Optimizing the ``distance'' of bridging information---designing ``gradual bridging'' that remains within agents' confidence thresholds while promoting expansion of information spheres---is an important research direction in conjunction with the findings of \citet{bail2018exposure}.

Fifth, the operationalization of ADI in Section~4.1 introduces a normalization convention ($G, I, F \in [0,1]$, $B \geq \varepsilon$) but does not resolve the deeper question of how each variable should be measured against a common empirical scale across historical stages. Making ADI a true measurement instrument rather than a comparative-descriptive index requires a systematic operational procedure that we leave to future empirical work.

Sixth, the $(V, S)$ values used in the simulation are illustrative representative endpoints, not empirical estimates. A robustness check with perturbed $(V, S)$ across a wider grid, together with reporting of variance across multiple random seeds, is a straightforward extension that would strengthen the empirical claims.

Seventh, the manipulation of $F$ was modeled in a simplified way as the precision of the algorithm's observation of interest vectors, but actual feedback resolution includes multiple dimensions abstracted away in this model, such as temporal delay, diversity of measurement targets (viewing time, clicks, sentiment analysis of comments, etc.), and upper limits on measurement precision due to privacy constraints. More refined modeling of $F$ that accounts for these factors is a task for future work.

Finally, a note on the relationship between descriptive analysis and normative claims. The descriptive claim of this paper is that echo chamber formation is an inevitable consequence of human cognitive constraints; this fact does not directly lead to the normative conclusion that ``echo chambers are desirable.'' The normative claim of this paper is a methodological position that, accepting cognitive constraints as given, we should discuss feasible institutional design. The normative question of whether echo chambers are desirable is outside the scope of this paper.

\section{Conclusion}
\label{sec:conclusion}

This paper constructed ADI, which quantitatively describes the decentralization of agenda-setting power, and the SIH model, which integrates the degree of decentralization with bridging strength, and verified the effectiveness of domain-selective bridging through agent-based simulation.

The starting point of the analysis is the recognition that echo chambers are an inevitable consequence of human evolutionary cognitive constraints and algorithmic engagement optimization. Based on this recognition, conventional prescriptions that normatively criticize echo chambers and demand individual behavioral change have low feasibility given the limits of human cognitive capacity. A realistic response is to incorporate not only engagement but also bridging into algorithmic scoring functions and to optimize the weights according to the characteristics of each information domain.

The simulation results demonstrated that domain-selective bridging is overwhelmingly superior to uniform bridging in simultaneously achieving social information health and user satisfaction. Uniform bridging carries the risk of collapsing user satisfaction and is not a sustainable design for either platform operators or users. Domain-selective bridging concentrates bridging resources on information domains where verifiability is low and collective scope is large, thereby improving information sharing in domains relevant to democratic decision-making without undermining user experience in hobby and lifestyle information.

The implication of this paper is that it liberates the echo chamber debate from the binary opposition of ``whether or not to eliminate echo chambers'' and transforms it into an engineering design problem amenable to treatment: ``in which information domains, to what degree, and through what algorithm design should bridging be realized?'' This transformation enables progress from a perpetual stalemate of normative debate to testable hypotheses and implementable design guidelines.

\subsection*{Funding}
This work was supported by JSPS KAKENHI Grant-in-Aid for Scientific Research (B) (Grant Number 26K00269).

\subsection*{AI Use Declaration}
This paper was drafted with the assistance of Claude (Anthropic), a large language model. The AI was used for literature review support, drafting assistance, simulation code development, and iterative revision. All intellectual contributions, theoretical arguments, conceptual frameworks, and final editorial decisions are the sole responsibility of the author.

\subsection*{Conflict of Interest}
The author declares no conflicts of interest. The author has no financial or professional relationships with any media organizations, platform companies, or technology firms discussed in this paper.

\bibliographystyle{apalike}
\bibliography{references}

@book{abernathy2018expanding,
  author    = {Abernathy, Penelope Muse},
  title     = {The Expanding News Desert},
  publisher = {Center for Innovation and Sustainability in Local Media, University of North Carolina at Chapel Hill},
  year      = {2018}
}

@book{anderson2006long,
  author    = {Anderson, Chris},
  title     = {The Long Tail: Why the Future of Business Is Selling Less of More},
  publisher = {Hyperion},
  year      = {2006}
}

@article{bail2018exposure,
  author  = {Bail, Christopher A. and Argyle, Lisa P. and Brown, Taylor W. and Bumpus, John P. and Chen, Haohan and Hunzaker, M. B. Fallin and Lee, Jaemin and Mann, Marcus and Merhout, Friedolin and Volfovsky, Alexander},
  title   = {Exposure to Opposing Views on Social Media Can Increase Political Polarization},
  journal = {Proceedings of the National Academy of Sciences},
  volume  = {115},
  number  = {37},
  pages   = {9216--9221},
  year    = {2018}
}

@article{bakshy2015exposure,
  author  = {Bakshy, Eytan and Messing, Solomon and Adamic, Lada A.},
  title   = {Exposure to Ideologically Diverse News and Opinion on {Facebook}},
  journal = {Science},
  volume  = {348},
  number  = {6239},
  pages   = {1130--1132},
  year    = {2015}
}

@book{chadwick2013hybrid,
  author    = {Chadwick, Andrew},
  title     = {The Hybrid Media System: Politics and Power},
  publisher = {Oxford University Press},
  year      = {2013}
}

@book{cohen1963press,
  author    = {Cohen, Bernard C.},
  title     = {The Press and Foreign Policy},
  publisher = {Princeton University Press},
  year      = {1963}
}

@article{deffuant2000mixing,
  author  = {Deffuant, Guillaume and Neau, David and Amblard, Fr\'{e}d\'{e}ric and Weisbuch, G\'{e}rard},
  title   = {Mixing Beliefs Among Interacting Agents},
  journal = {Advances in Complex Systems},
  volume  = {3},
  number  = {01n04},
  pages   = {87--98},
  year    = {2000}
}

@inproceedings{donkers2021dual,
  author    = {Donkers, Tim and Ziegler, J\"{u}rgen},
  title     = {The Dual Echo Chamber: Modeling Social Media Polarization for Interventional Use},
  booktitle = {Proceedings of the 2021 Workshop on Open Challenges in Online Social Networks},
  year      = {2021}
}

@article{dubois2018echo,
  author  = {Dubois, Elizabeth and Blank, Grant},
  title   = {The Echo Chamber Is Overstated: The Moderating Effect of Political Interest and Diverse Media},
  journal = {Information, Communication \& Society},
  volume  = {21},
  number  = {5},
  pages   = {729--745},
  year    = {2018}
}

@article{dunbar1992neocortex,
  author  = {Dunbar, Robin I. M.},
  title   = {Neocortex Size as a Constraint on Group Size in Primates},
  journal = {Journal of Human Evolution},
  volume  = {22},
  number  = {6},
  pages   = {469--493},
  year    = {1992}
}

@article{gentzkow2011ideological,
  author  = {Gentzkow, Matthew and Shapiro, Jesse M.},
  title   = {Ideological Segregation Online and Offline},
  journal = {Quarterly Journal of Economics},
  volume  = {126},
  number  = {4},
  pages   = {1799--1839},
  year    = {2011}
}

@article{granovetter1973strength,
  author  = {Granovetter, Mark S.},
  title   = {The Strength of Weak Ties},
  journal = {American Journal of Sociology},
  volume  = {78},
  number  = {6},
  pages   = {1360--1380},
  year    = {1973}
}

@article{guess2023algorithms,
  author  = {Guess, Andrew M. and Malhotra, Neil and Pan, Jennifer and Barber\'{a}, Pablo and Allcott, Hunt and Brown, Taylor and others},
  title   = {How Do Social Media Feed Algorithms Affect Attitudes and Behavior in an Election Campaign?},
  journal = {Science},
  volume  = {381},
  number  = {6656},
  pages   = {398--404},
  year    = {2023}
}

@article{guess2023reshares,
  author  = {Guess, Andrew M. and Malhotra, Neil and Pan, Jennifer and Barber\'{a}, Pablo and Allcott, Hunt and Brown, Taylor and others},
  title   = {Reshares on Social Media Amplify Political News but Do Not Detectably Affect Beliefs or Opinions},
  journal = {Science},
  volume  = {381},
  number  = {6656},
  pages   = {404--408},
  year    = {2023}
}

@book{guo2016power,
  author    = {Guo, Lei and McCombs, Maxwell},
  title     = {The Power of Information Networks: New Directions for Agenda Setting},
  publisher = {Routledge},
  year      = {2016}
}

@article{hegselmann2002opinion,
  author  = {Hegselmann, Rainer and Krause, Ulrich},
  title   = {Opinion Dynamics and Bounded Confidence: Models, Analysis and Simulation},
  journal = {Journal of Artificial Societies and Social Simulation},
  volume  = {5},
  number  = {3},
  year    = {2002}
}

@book{mccombs2014setting,
  author    = {McCombs, Maxwell E.},
  title     = {Setting the Agenda: Mass Media and Public Opinion},
  edition   = {2nd},
  publisher = {Polity Press},
  year      = {2014}
}

@article{mccombs1972agenda,
  author  = {McCombs, Maxwell E. and Shaw, Donald L.},
  title   = {The Agenda-Setting Function of Mass Media},
  journal = {Public Opinion Quarterly},
  volume  = {36},
  number  = {2},
  pages   = {176--187},
  year    = {1972}
}

@article{meraz2009elite,
  author  = {Meraz, Sharon},
  title   = {Is There an Elite Hold? {Traditional} Media to Social Media Agenda Setting Influence in Blog Networks},
  journal = {Journal of Computer-Mediated Communication},
  volume  = {14},
  number  = {3},
  pages   = {682--707},
  year    = {2009}
}

@article{nyhan2023likeminded,
  author  = {Nyhan, Brendan and Settle, Jaime and Thorson, Emily and Wojcieszak, Magdalena and Barber\'{a}, Pablo and Chen, Annie Y. and others},
  title   = {Like-Minded Sources on {Facebook} Are Prevalent but Not Polarizing},
  journal = {Nature},
  volume  = {620},
  number  = {7972},
  pages   = {137--144},
  year    = {2023}
}

@techreport{ovadya2023bridging,
  author      = {Ovadya, Aviv and Thorburn, Luke},
  title       = {Bridging Systems: Open Problems for Countering Destructive Divisiveness Across Ranking, Current Events, and Dialog},
  institution = {Knight First Amendment Institute, Columbia University},
  year        = {2023}
}

@book{pariser2011filter,
  author    = {Pariser, Eli},
  title     = {The Filter Bubble: What the Internet Is Hiding from You},
  publisher = {Viking/Penguin},
  year      = {2011}
}

@book{prior2007post,
  author    = {Prior, Markus},
  title     = {Post-Broadcast Democracy: How Media Choice Increases Inequality in Political Involvement and Polarizes Elections},
  publisher = {Cambridge University Press},
  year      = {2007}
}

@book{shoemaker2014mediating,
  author    = {Shoemaker, Pamela J. and Reese, Stephen D.},
  title     = {Mediating the Message in the 21st Century: A Media Sociology Perspective},
  edition   = {3rd},
  publisher = {Routledge},
  year      = {2014}
}

@book{sperber1995relevance,
  author    = {Sperber, Dan and Wilson, Deirdre},
  title     = {Relevance: Communication and Cognition},
  edition   = {2nd},
  publisher = {Blackwell},
  year      = {1995},
  note      = {Original work published 1986}
}

@book{starr2004creation,
  author    = {Starr, Paul},
  title     = {The Creation of the Media: Political Origins of Modern Communications},
  publisher = {Basic Books},
  year      = {2004}
}

@book{sunstein2001republic,
  author    = {Sunstein, Cass R.},
  title     = {Republic.com},
  publisher = {Princeton University Press},
  year      = {2001}
}

@book{sunstein2017republic,
  author    = {Sunstein, Cass R.},
  title     = {\#Republic: Divided Democracy in the Age of Social Media},
  publisher = {Princeton University Press},
  year      = {2017}
}

@incollection{tooby1992psychological,
  author    = {Tooby, John and Cosmides, Leda},
  title     = {The Psychological Foundations of Culture},
  booktitle = {The Adapted Mind: Evolutionary Psychology and the Generation of Culture},
  editor    = {Barkow, Jerome H. and Cosmides, Leda and Tooby, John},
  pages     = {19--136},
  publisher = {Oxford University Press},
  year      = {1992}
}

@article{vargo2014network,
  author  = {Vargo, Chris J. and Guo, Lei and McCombs, Maxwell and Shaw, Donald L.},
  title   = {Network Issue Agendas on {Twitter} During the 2012 {U.S.} Presidential Election},
  journal = {Journal of Communication},
  volume  = {64},
  number  = {2},
  pages   = {296--316},
  year    = {2014}
}

@book{williams2011after,
  author    = {Williams, Bruce A. and Delli Carpini, Michael X.},
  title     = {After Broadcast News: Media Regimes, Democracy, and the New Information Environment},
  publisher = {Cambridge University Press},
  year      = {2011}
}

\end{document}